\documentclass[twocolumn,showpacs,prl,amsmath,amssymb,epsfig]{revtex4}

\usepackage{graphicx}% Include figure files
\usepackage{dcolumn}% Align table columns on decimal point
\usepackage{bm}% bold math

\begin{document}

\title{Possible evidence of a spontaneous spin-polarization in
mesoscopic 2D electron systems}
\author{A. Ghosh, C. J. B. Ford, M. Pepper, H. E. Beere, D. A. Ritchie}
\affiliation{Cavendish Laboratory, University of Cambridge,
Madingley Road, Cambridge CB3 0HE, United Kingdom}
\date{\today}

\begin{abstract}
We have experimentally studied the non-equilibrium transport in
low-density clean 2D electron systems at mesoscopic length scales.
At zero magnetic field ($B$), a double-peak structure in the
non-linear conductance was observed close to the Fermi energy in
the localized regime. From the behavior of these peaks at non-zero
$B$, we could associate them to the opposite spin states of the
system, indicating a spontaneous spin polarization at $B = 0$.
Detailed temperature and disorder dependence of the structure
shows that such a splitting is a ground state property of the
low-density 2D systems.
\end{abstract}

 \maketitle

Spin polarization of electrons in low dimensional systems at $B =
0$ has recently attracted extensive theoretical and experimental
attention. In the metallic regime, this is commonly attributed to
the spin-orbit (SO) effect, where the bulk inversion asymmetry
(the $k^3$ term) and the interface inversion asymmetry (the Rasba
term) result in an energy separation of several millivolts between
the spin bands in narrow band semiconductors~\cite{das}. At low
electron densities however, the electron-electron interaction
dominates over the kinetic energy, and a homogeneous 2D system
becomes unstable to spontaneous spin polarization (SSP) due to
exchange~\cite{varsano,dharma}. Since the influence of interaction
is strong at lower dimensions, several recent experimental
investigations on the spin state of 0D~\cite{stewart,folk} and 1D
systems~\cite{thomas} have indicated the possibility of a SSP,
even though the origin and nature of such a phase is
controversial. In low-density 2D electron systems (2DES's),
however, a direct observation of a ``spin gap'' has remained
experimentally illusive, even though evidence of an enhanced
g-factor and anomalous spin-susceptibility have been reported when
the disorder is low~\cite{tutuc,okamoto,shashkin,zhu}.

It is now known that, an unpaired spin in weakly-isolated systems
are screened by Kondo-like singlet formation with the electrons in
the leads~\cite{kondot}. This results in a resonance in the
tunnelling density-of-states (DOS), and an enhancement in the
differential conductance ($dI/dV$) through the system when the
chemical potential of the leads are aligned. When studied as a
function of source-drain bias ($V_{SD}$), the zero-bias peak (ZBP)
splits linearly in finite $B$ by $\Delta_Z = g^*\mu_BB$, where
$g^*$ and $\mu_B$ are the effective g-factor and the Bohr magneton
respectively. Recent non-linear studies in ballistic quantum point
contacts (QPC's) have also shown the evidence of a
ZBP~\cite{kond1d}, which was attributed to a Kondo-like
correlation resulting from a dynamic SSP in 1D. In this Letter, we
report the direct observation of SSP in low-density 2DES's. High
quality 2DES's of mesoscopic length scales were used with no
intentional in-plane confinements. In the localized regime, clear
evidence of a split ZBP was observed at $B = 0$. The magnitude of
the split ($\Delta$) evolved continuously with $B$, implying its
origin to be related to the underlying spin structure of the 2DES.
Simultaneous measurement of the Fermi energy ($E_F$) showed that
such a polarization is an intrinsic ground state property of 2D
systems, and depends critically on the impurity scattering, as
well as the temperature ($T$) and magnetic field ($B$).

We have used 2DES's confined to the triangular potential wells of
Si $\delta$-doped GaAs/AlGaAs heterostructures. In most samples,
the 2DES's were formed $\approx$ 300 nm below the surface.
Disorder was varied by changing the spacer thickness $\delta_{sp}$
separating the dopant layer from the GaAs/AlGaAs interface. Data
from two samples with $\delta_{sp}$ = 40 nm and 60 nm (referred as
A78 and A79, respectively) are reported in this work, even though
all the samples show qualitatively similar results. In both
samples, a source-drain voltage $V_{SD}$ was applied on a
$5\mu$m$\times5\mu$m region of the wafer, defined by an etched
mesa and a metallic surface gate. At zero gate voltage ($V_g = 0$)
the mobility of both samples were $\gtrsim 2\times10^6$
cm$^2$/V-sec. Lifetime measurements confirm the dominance of the
small angle scattering. By varying $V_g$, electron density ($n_s$)
as low as $\sim 5\times10^9$ cm$^{-2}$ could be attained in sample
A79 (corresponding to an interaction parameter $r_s =
1/a_B^*\sqrt{\pi n_s} \sim 7.6$, where $a_B^*$ is the effective
Bohr radius). In all the magnetic field measurements, $B$ was
applied in the plane of the 2DES and parallel to the direction of
the current. The differential conductance $dI/dV$ was measured
with a standard 2-probe mixed ac-dc method, where the ac
excitation bias was kept at $< k_BT/e$. For all samples, the gate
was trained several hundred times to obtain an excellent run to
run reproducibility (better than 0.1\%).

Dependence of the linear response conductance ($G$) on $V_g$ is
shown in Fig.~1 at various values of $B$, recorded at $T \approx
35$ mK in sample A79. We focus on the localized regime (MIT in A79
occurs at $G \sim 3\times e^2/h$), where the electron density is
low, and the interaction effects are most pronounced. For all
$V_g$, $G$ was found to decrease rapidly with increasing $B$. The
functional dependence of the in plane magneto-conductance (MC) on
$B$ has been shown to change from $\sim e^{-B^2}$ to $\sim e^{-B}$
at a critical field $B_c$, when the electron gas becomes
completely spin polarized, and the majority and minority spin
bands are separated by the Fermi energy~\cite{tutuc,zhu}. In the
inset of Fig.~1 we have shown the zero-bias MC as a function of
$B^2$ at four representative gate voltages. The values of $B_c$ at
each $V_g$, shown by the arrows, indicate a deviation from the
linear behavior at low fields. This enables us to evaluate the
spin non-degenerate Fermi energy $E_F^*$ as $E_F^* = g^*\mu_BB_c$.
The dependence of $E_F^*$ on $V_g$ is shown in Fig.~2c. We have
used a $g^* = 3.4|g_b|$ for evaluating $E_F^*$, where $|g_b| =
0.44$ is the band g-factor in bulk GaAs. This value of $g^*$ is
obtained from $V_{SD}$ measurements, and will be discussed later.
Note that above $V_g \approx -0.376$ V, corresponding to $n_s
\approx 5\times10^9$ cm$^{-2}$, the transport is essentially 2D in
nature with $E_F^*$ varying approximately linearly with $V_g$. The
slope $dE_F^*/dV_g$ was found to be $\approx 12$ meV/V, agreeing
roughly with the effective free-electron spin non-degenerate 2D
density-of-states, $(dE_F^*/dV_g)_{free} = (h^2/2\pi m^*)dn_s/dV_g
\approx \epsilon_0\epsilon_r h^2/2\pi em^*d_s \approx 16$ meV/V,
where $d_s = 310$ nm is the depth of the 2DES from the surface,
and $m^* \approx 0.067m_e$ is the band effective mass of the
electron. The discrepancy could be due to a weak
density-dependence of $g^*$~\cite{tutuc}. Below $V_g \approx
-0.376$ V ($G \approx 0.3-0.4\times e^2/h$), $E_F^*$ drops
abruptly, possibly due to the onset of inhomogeneity in the charge
distribution as screening becomes weak.

The $V_{SD}$ dependence of $dI/dV$ at various values of $V_g$
(i.e., $n_s$) is shown in Fig.~2a. $V_g$ differs by 1 mV in
successive offset-corrected traces. The striking feature of these
traces is the double peak structure of $dI/dV$ with a local
minimum at $V_{SD} = 0$. This was found to be a generic feature
observed in all the low-disorder samples of similar dimensions.
The detection of the double-peak structure was difficult in the
strongly localized regime ($V_g \lesssim -0.385$ V), as well as in
the metallic regime ($V_g \gtrsim -0.35$ V), indicating a
disorder-dependent window of $n_s$ where the effect becomes
clearly visible. In most cases the peaks are dissimilar in
magnitude and width, both of which vary when $V_g$ is changed. The
separation ($\Delta$) of the peaks shows a non-monotonic
dependence on $V_g$, as shown in Fig.~2b. At low $V_g$ ($\sim
-0.385$ V), $\Delta$ is largest, but decreases rapidly with
increasing $V_g$, reaching a minimum at $V_g \approx -0.375$ V.
Comparing to the $V_g$ dependence of $E_F^*$ (Fig.~2c) we find the
onset of linear dependence of $E_F^*$ at the same $V_g$. When $V_g
\gtrsim -0.375$ V, $\Delta$ increases roughly linearly with
increasing $V_g$. Since this regime can be directly associated
with a 2D ground state, we shall restrict further discussions on
$\Delta$ to this range of $V_g$. Extrapolating the linear
dependence of $\Delta$, we find the $V_g$ ($\approx -0.395$ V) at
which $\Delta = 0$, agrees within the experimental uncertainty
with the $V_g$ ($\approx -0.408$ V) at which the extrapolated
$E_F^* = 0$, establishing a direct correspondence between $\Delta$
and $n_s$.

In order to investigate the nature of this effect, we have then
studied the double peak structure in a parallel magnetic field.
The behavior is illustrated with the trace measured at $V_g
\approx -0.37$ V, indicated by the arrow in Fig.~2b. This is shown
in Fig.~3, where the traces with increasing $B$ are vertically
offset for clarity. The peak positions ($V_p$) change
non-monotonically as a function of $B$ (see the arrows). We find
that the peaks close in over the field scale of $B \lesssim 0.5$
T, but eventually separate at higher fields (see the inset of
Fig.~3). The evolution of the peaks with $B$ strongly suggests a
spin-related effect, where $\Delta$ is the energy difference
between the opposite spins. Note that the $B$-field does not split
each peak individually, differentiating our case from a disorder
induced quantum molecule. An asymmetrical suppression of the peaks
with increasing $B$ sets the maximum field scale ($\approx 2.8$ T)
of our experiments, beyond which the left peak becomes
undetectable.

The observability and characteristics of the peaks were found to
depend critically on disorder and $T$. We have changed the local
disorder profile over the sample region by controlled thermal
cycles from room temperature to 4.2 K. The result for three
successive thermal cycles carried out for A78 are shown in
Fig.~4a. All traces (B and C are shifted vertically for clarity)
were recorded at a similar $E_F$ and $B = 0$. While the general
behavior of $dI/dV$ is similar and agrees with that of A79, we
find the overall magnitude and width of the peaks to differ
markedly, even from one cooling to the other. In general, when the
cooling was done at a slow rate (trace A: over a few hours) we
found the peaks to be more pronounced than when the cooling was
done rapidly (trace C: over few tens of minutes). Greater disorder
in C is also observable in terms of the linear response
conductance, $G \approx 0.1e^2/h$, which, in case of A is $\approx
0.3e^2/h$. Disorder broadening also affects the observability of
the effect as the sample dimensions are increased. Typically, no
double-peaks were observed in $10\mu$m$\times10\mu$m samples.

Thermal broadening of the peaks, associated with a strong
suppression of the peak height, from $T \sim 70$ mK to $\sim 0.5$
K are shown in Fig.~4b. Note that even though $G$ at $V_{SD} = 0$
rises with increasing $T$, as expected in a localized system, it
is essentially a result of the overlap of the broadening peaks.
This also confirms that as a function of $V_{SD}$, we indeed
observe a non-linear effect, and the double peak structure is a
ground-state property of the 2D electron system in the low density
regime.

We now discuss possible mechanisms which may give rise to such a
structure in non-equilibrium measurements. Assuming the formation
of a weak tunnel barrier, which physically splits the 2D region in
two parts, the conservation of the transverse momentum will allow
tunnelling through the barrier only at $V_{SD} = 0$. This would
result in a disorder-broadened ZBP in the tunnelling $dI/dV$. An
interaction-induced suppression of states at $E_F$, e.g., $\sim
\ln{(|\epsilon|\tau)}$ in the diffusive regime~\cite{altshuler} or
Efros-Shklovskii-type soft gap $\sim |\epsilon|$ in the hopping
regime~\cite{es}, may then spilt the ZBP to give rise to the
double peak structure~\cite{rudin}. The main arguments against
such a scenario are, (a) the separation of the peaks are
$B$-independent, hence inconsistent with Fig.~3, and (b) the
separation decreases with increasing $E_F$, contrary to the result
of Fig.~2b~\cite{rudin}.

To investigate if the peak separation $\Delta$ is indeed
spin-related, we have plotted $\Delta$ as a function of $B$ for
the data shown in fig.~3. As shown in Fig.~5, at low fields
$\Delta$ decreases with increasing field, while at high $B$
($\gtrsim 1$ T), $\Delta$ shows monotonic increase as $B$
increases. In this high-$B$ regime, $\Delta$ tends asymptotically
to a linear $B$ dependence (the solid line), which when
extrapolated, passes through the origin. This we identify as
Zeeman spin splitting. From the asymptote, $\Delta_Z = g^*\mu_BB$,
we find $g^*$ to be $\approx 1.49 = 3.4|g_b|$, agreeing closely to
the values reported in recent measurements over similar range of
$n_s$~\cite{tutuc}. The peak positions thus represent the energies
of two opposite spin states in a magnetic field, and from the
non-zero separation at $B = 0$, signify a non-zero spin
polarization.

From the critical role of spin , and also the behavior of $dI/dV$
near $V_{SD} = 0$, a Kondo-like many-body correlation, as
discussed extensively in the context of quantum
dots~\cite{kondot}, can be envisaged. There are however important
differences. In our system, as in the case of QPC's, there are no
obvious singly-occupied quasi-bound electronic states. Unlike the
suggested ferromagnetic states in QPC's~\cite{karl}, the magnetic
moment of a frozen spin-polarized state in 2D would be much higher
than that of a single electron, and hence Kondo-type screening
would be difficult. However, as indicated in Fig.~2a for $V_g
\gtrsim -0.375$ V, the spin-polarization in our system seems to be
dynamic, and increases with $n_s$. This, as suggested for
QPC's~\cite{meir}, in presence of optimal band-hybridization,
could result in a Kondo-enhanced ZBP in $dI/dV$. Our case is
however further complicated by the splitting of the ZBP at $B =
0$, resembling the behavior of ZBP in dots and QPC's at a finite
$B$~\cite{kondot,kond1d}. Although the Kondo effect in coupled dot
systems has a similar non-equilibrium behavior as a function of
$V_{SD}$ at $B = 0$, such a case would appear only over a
restricted parameter range of inter-dot and lead couplings, and
hence should be rather rare in open systems with no intentional
confinement~\cite{ddot}.

Finally, in parallel $B$, it has been shown that in sufficiently
smooth disorder, both SO and exchange-induced spin splitting could
result in satellite-peaks in the tunnelling DOS at finite bias
$\approx \pm \sqrt{\Delta_0^2 + \Delta_Z^2 }$, where $\Delta_0$ is
the magnitude of SSP~\cite{apalkov}. The tunnelling DOS could be
obtained in a non-linear $dI/dV$ measurement if we assume a
quasi-ballistic transport in our clean mesoscopic samples, i.e.,
electron energies are lost mainly at the leads. Apart from
explaining the double-peak structure in $dI/dV$, this also
justifies, (a) broadening of the peaks with disorder and (b) the
linear $B$-dependence of $\Delta$ at high $B$, where $\Delta
\rightarrow \Delta_Z$. If $\Delta_0$ arises from a strong SO
coupling, one expects negative magneto-conductance at low
perpendicular $B$-field, arising from weak antilocalization. No
evidence of antilocalization was observed in our mesoscopic 2D
systems. The SO-origin of $\Delta_0$ seems to be unlikely on two
more grounds. First, calculation of the magnitude of splitting in
GaAs/AlGaAs heterostructures, taking into account both bulk and
Rashba terms, shows that $\Delta_{0}$ would be $\ll 0.05$ meV
below $n_s \sim 10^{11}$ cm$^{-2}$~\cite{pfeffer}. Secondly, if
the absence of antilocalization  is attributed to a $B$-dependent
$\Delta_0$, as in large open quantum dots, $\Delta_0$ increases
with $B$ at low fields~\cite{folk,halperin}. However, as shown in
the inset of Fig.~5, $\Delta_0 \sim \sqrt{\Delta^2 - \Delta_Z^2}$
decreases with increasing $B$ in our samples. Alternatively, if we
assume that $\Delta_0$ arises due to exchange, and in the Fermi
liquid limit, represents the shift of one spin state with respect
to the other (see the inset of Fig.~5), then an estimate of
polarization $\zeta$ could be obtained as $\zeta = (n_{s\uparrow}
- n_{s\downarrow})/n_s \sim \Delta/2E_F^*$. In the regime of 2D
conduction, $V_g \gtrsim -0.375$ V, we find $\zeta \sim 0.1$,
implying only a partial spin polarization. This is probably not
surprising considering the low $r_s \sim 6 - 7$ in our system, and
finite $T$~\cite{dharma}. The exchange-origin of $\Delta_0$ is
also supported by the decrease of $\Delta_0$ with increasing $B$,
as the additional confinement ($\sim B^2$) imposed by $B$ tends to
counteract the formation of parallel spins~\cite{karl}.

In conclusion, non-linear conductance measurements in high quality
mesoscopic 2D electron systems show direct evidence of spontaneous
spin polarization at low electron densities. This spin
polarization is density-dependent, and manifested by a split
zero-bias peak in source-drain bias at $B = 0$. The peaks are
generally asymmetric and strongly depend on the local disorder.
Simultaneous measurement of the Fermi energy, and
temperature-dependence of conductance confirm the splitting to be
an intrinsic property of 2D systems, and also identify a window of
electron density and temperature over which the splitting becomes
experimentally observable.

\begin{figure*}[h]
\centering
\includegraphics[height=10cm,width=13cm]{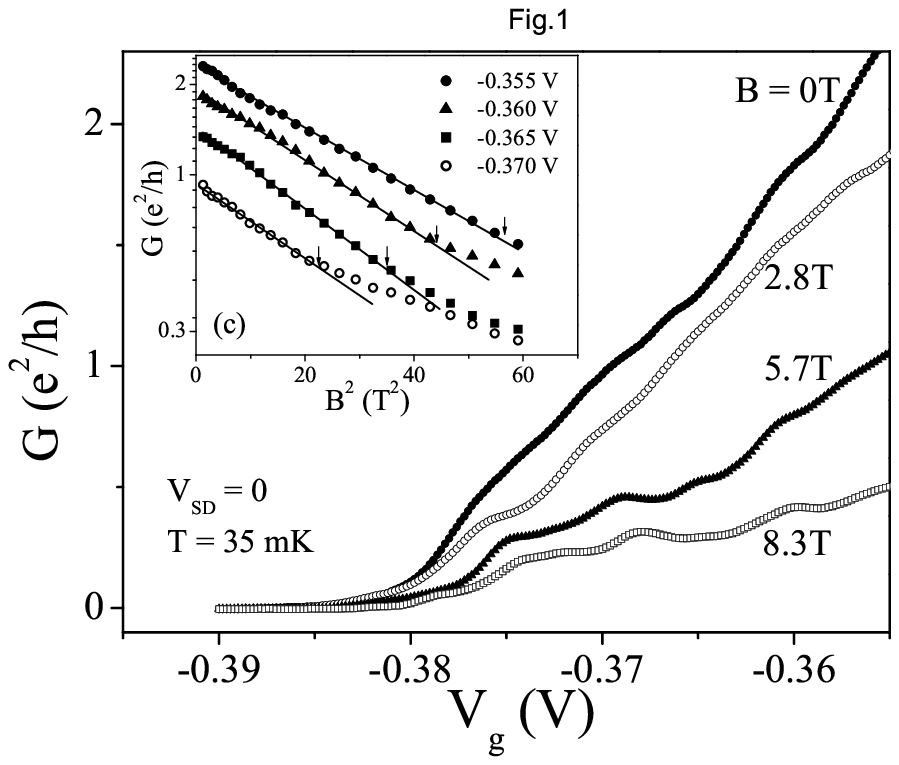}
\caption{Typical gate voltage $V_g$ dependence of the linear
response conductance $G$ at various magnetic fields. Inset: $G$,
as a function of $B^2$ at four $V_g$'s. Arrows indicate the
deviation from linear dependence, and the onset of complete
polarization.}
\end{figure*}
\begin{figure*}[h]
\centering
\includegraphics[height=10cm,width=13cm]{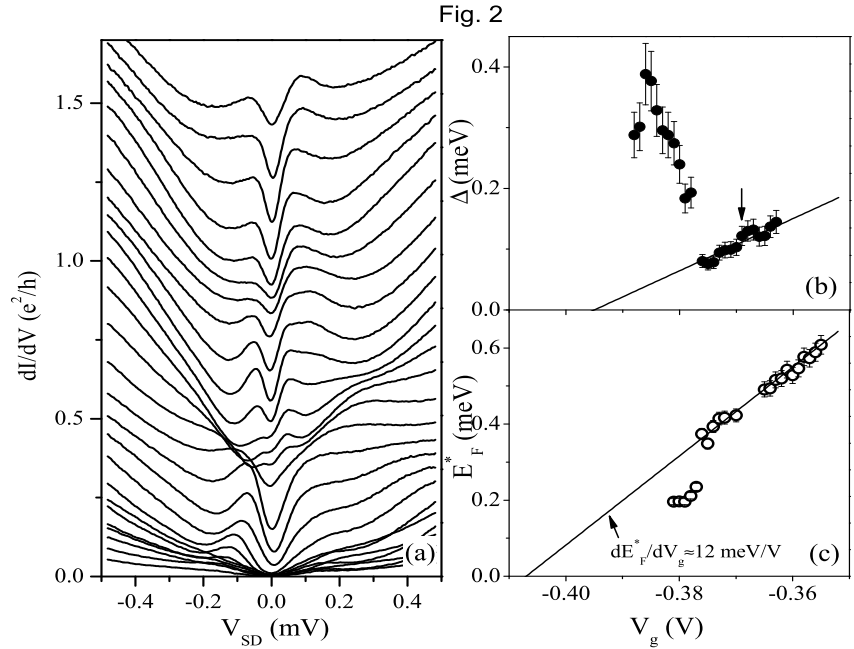}
\caption{(a) The differential conductance $dI/dV$ vs. source-drain
bias $V_{SD}$ at various $V_g$. $V_g$ differs by 1 mV for
successive sweeps. (b) The peak separation $\Delta$ as a function
of $V_g$. (c) $V_g$-dependence of the spin non-degenerate Fermi
energy $E_F^*$, obtained from the magnetic field scale of complete
field polarization (see text).}
\end{figure*}
\begin{figure*}[h]
\centering
\includegraphics[height=10cm,width=13cm]{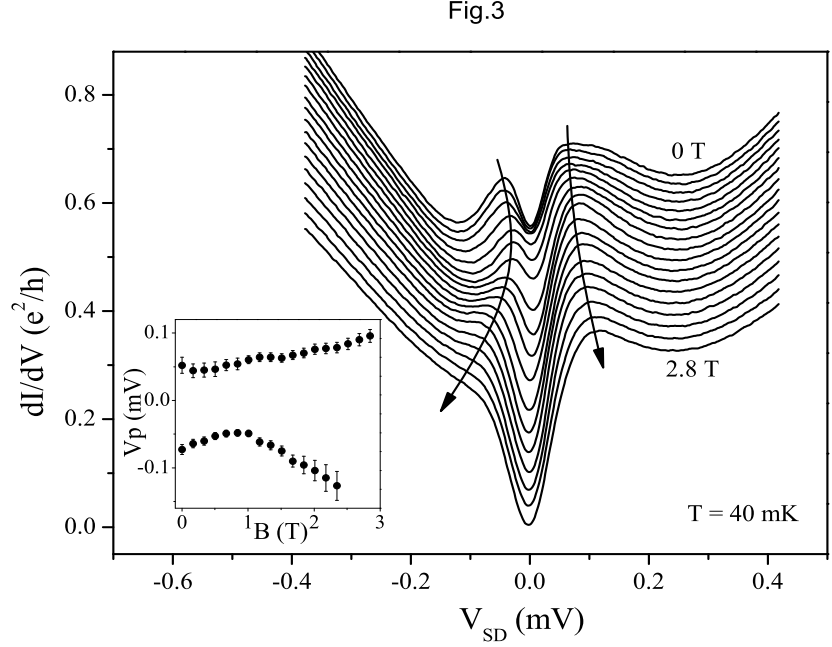}
\caption{Evolution of the differential conductance with magnetic
field $B$. We have illustrated this with the trace obtained at
$V_g \approx -0.37$ V (indicated by the arrow in Fig.~2b). Traces
are shifted vertically for clarity. Inset: Positions ($V_p$) of
the peaks as a function of $B$.}
\end{figure*}
\begin{figure*}[h]
\centering
\includegraphics[height=10cm,width=13cm]{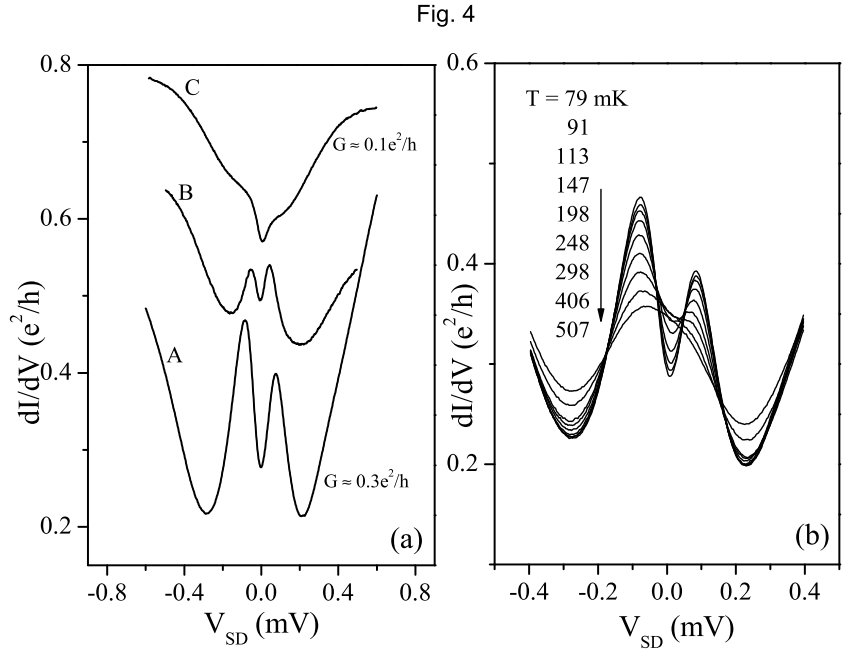}
\caption{(a) Double-peak structure in $dI/dV$ in the same sample
for various cool-downs. Traces are shifted vertically for easy
comparison. The linear response conductance is minimum for C and
maximum for A. (b) Temperature dependence of the double peak
structure. Note the smearing of each peak results in the increase
of the zero-bias conductance with increasing $T$.}
\end{figure*}
\begin{figure*}[h]
\centering
\includegraphics[height=10cm,width=13cm]{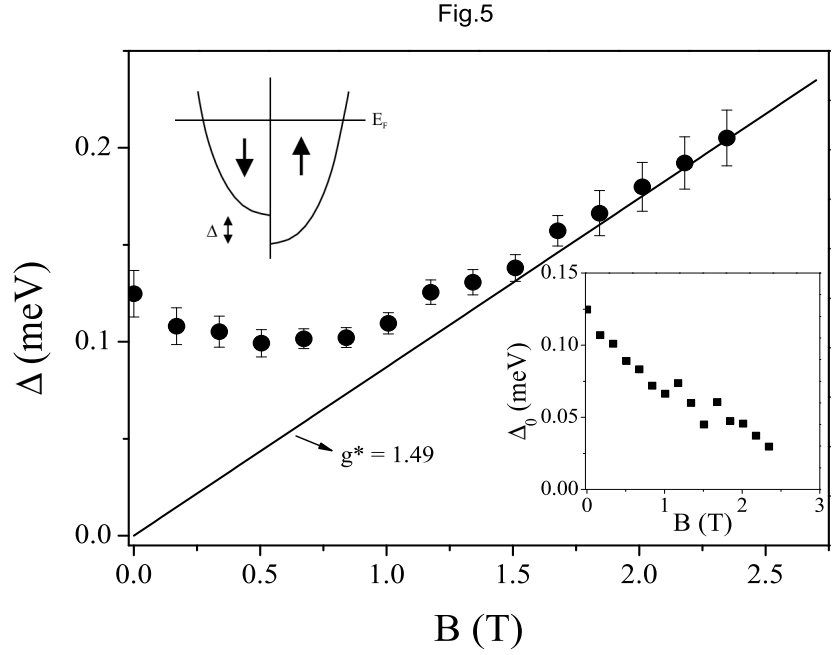}
\caption{Dependence of the peak separation $\Delta$ on the
magnetic field $B$. We have used the $\Delta$ shown in Fig.~3. The
high-field asymptote extrapolates through the origin. Inset:
$B$-dependence of the $\Delta_0$.}
\end{figure*}

\end{document}